# Development of a 'Game with a Purpose' for Acquisition of Brain-Computer Interface Data

Joe T. Rexwinkle[1], Gregory Lieberman[1], Matthew Jaswa[2], Brent J. Lance[1]

[1]CCDC U.S. Army Research Laboratory, Human Research and Engineering Directorate
[2]DCS Corporation

**Corresponding Author:** Joe Rexwinkle (joe.t.rexwinkle.civ@mail.mil)

**Abstract**

Brain-computer interfaces (BCIs) have the potential to significantly change the ways in which humans interact with technology, the environment, and even each other. Unfortunately, BCI technologies are seldom robust enough for use in real-world applications, in part due to the large amount of data that must be collected, processed, and classified in order to develop models of task-related neural activity that account for two of the most important and least-understood drivers of BCI illiteracy: individual differences in neural signals and intra-individual differences across interdependent, time-varying neural states. This paper describes the feasibility of using a game with a purpose (GWAP) as a viable instrument for collecting data from BCI-relevant research tasks. By leveraging game-related reward processes to maintain participant interest and engagement, this approach will enable large amounts of BCI data to be acquired, both across many individuals and longitudinally from specific individuals as neural states vary naturally over time. Pilot and technical testing results are presented here to demonstrate that the BCI-relevant tasks embedded within the research game elicit neural signals similar to those that would be expected from more traditional BCI tasks. These preliminary data provide support and validation of the use of GWAPs as promising tools to enable long-term collection of BCI-relevant data in an engaging environment.

## 1. Introduction

Brain-computer interface (BCI) technologies have the potential to greatly improve the interactions between humans and machines, the environment, and even other humans (Lance et al., 2012). While some BCIs leverage implanted cranial electrodes, the vast majority of these devices interpret neural signals collected using non-invasive electroencephalography (EEG) as the input for interfaces between humans and computer systems. However, due to volume conduction across the scalp and motion-induced (and other) artifacts, EEG signals tend to be noisy, making it difficult to model and interpret the underlying brain activity. Target phenomena are often isolated using frequency and/or spatial filters, but concurrent cognitive processes and mental states may interact in complex ways that make it difficult to filter out relevant target signals (Blankertz et al., 2008, 2011; Lawhern et al., 2018; Lotte and Guan, 2011).

Due to this inability to consistently filter EEG signals that are essentially irrelevant to a given task, BCIs have been shown to have unstable performance over long periods of time (Shenoy et al., 2006). Changes in the "brain states" of an individual, caused both by known qualitative conditions such as stress, sleep pressure, and fatigue, as well as underlying processes that may not be behaviorally defined are all hypothesized to be significant contributors to the inconsistency in classification accuracy of BCI-relevant brain signals (Curran and Stokes, 2003; Millán et al., 2010; Myrden and Chau, 2015; Shenoy et al., 2006). Prior studies have been conducted using EEG data to detect individuals' qualitative states, including drowsiness (Hsu and Jung, 2017; Wu et al., 2017), stress (Berka et al., 2008), and attentiveness (Hsu and Jung, 2017; Jung et al., 1997; Myrden and Chau, 2015), but no studies to date have investigated the effects of interactions between many states over time on BCI robustness. Further studies have attempted to characterize the effect that these states have upon the presentation of BCI-relevant EEG, though robust, longitudinal studies are not observed in the literature (Berka et al., 2008; Kong et al., 2014; Myrden and Chau, 2015; Ries et al., 2016; Xie et al., 2016). This lack of longitudinal data prevents the differentiation of specific brain states and the interactions between them, many of which may operate on various time scales (within-trial, within-session, weeks, months, etc.), and applying this knowledge to the development of more robust methods for classification of BCI-relevant EEG signals. One factor that may contribute to the lack of sufficient longitudinal data is the very nature of

BCI tasks, themselves, which can often be long, arduous, and even potentially tedious. As a result, longitudinal BCI studies are susceptible to high levels of participant dropout or distraction, and novel methods are needed in order to enable collection of sufficiently large data sets to account for both individual differences in neural signals and intra-individual differences that arise from changes in interdependent, time-varying "brain states."

One potentially novel method for collection of longitudinal BCI data is the game with a purpose (GWAP), specifically designed to enhance user experience while performing BCI-relevant tasks (Lance et al., 2015). The GWAP model has been demonstrated to be a valuable tool for encouraging individuals to participate in non-BCI research activities (Cooper et al., 2010; von Ahn, 2006; von Ahn and Dabbish, 2008), though no BCI-specific GWAP has been developed to date. An additional benefit to this framework is that increased motivation has been demonstrated to result in higher-quality event-related potential (ERP) and sensorimotor rhythm (SMR) responses, meaning that the more fun and varied experience provided through gamification of research tasks may not only increase extrinsic and intrinsic reward, which in turn may lead to better retention of participants, but may also result in higher-quality data (Kleih et al., 2010; Nijboer et al., 2008, 2010). This proof-of-concept study explores the use of an engaging tile-matching game that incorporates BCI-relevant tasks as mini-games that provide participants opportunities to earn power-ups. This game will be used for longitudinal assessment of the effects of natural variability and induced changes in participant brain states on patterns in EEG and other physiologic signals, which may lead to the development of novel and increasingly robust BCI algorithms and a stronger understanding of BCI literacy.

## 2. Materials and Methods

The novel platform for data collection presented here is a tile-matching game in which participants control a robot fighting against a progression of enemy robots. Players must swap gem tiles to/from anywhere on the screen to create a row or column of at least three gems of the same type, which will then disappear, deal damage to the enemy, and award points to the participant. Matching longer rows/columns or causing cascades of gem matches earn participants more points and deal increased damage to the enemy. After a set number of moves, or in reaction to certain participant-initiated mistakes, the enemy robot will attack. Throughout the game, BCI-relevant tasks are presented as mini-games that allow participants to earn power-ups that improve their ability to attack or defend against the enemy robot. The inclusion of leaderboards and medals (awarded based on past performance) reward participants for their performance, adding extra incentive to engage with the game and continue to provide high-quality behavioral and EEG data.

Data presented in this paper were collected both during internal technical testing of the game and during IRB-approved pilot sessions. Each pilot participant took part in a single gameplay session. At the beginning of each game session, participants were outfitted with an ABM B-Alert X24t EEG System headset (Advanced Brain Monitoring, Carlsbad, CA), which has 20 EEG electrodes in a standard 10-20 montage, two reference electrodes, and two extra electrodes used to collect electrocardiography (ECG) data. The reference electrodes were attached to the mastoids and the ECG electrodes were attached to the collarbones following manufacturer instructions. An impedance measurement and subsequent corrective action was performed at the start of each session using ABM software to help ensure data quality. Other contextual and physiologic data such as game performance, pupillometry, eye-tracking, and questionnaires regarding stress, sleep pressure, fatigue, and other states/traits were also collected, but analysis of these measures will be presented in a future paper. Lab Streaming Layer (LSL) was embedded in the game and used to synchronize data recording from various sources.

The game contains four play modes: normal, time-limited, move-limited, and "shot clock." The initial three modes (normal, time-limited, and move-limited) are present to further engage participants by providing constraints on gameplay and varied objectives: normal mode ends when the participant's or enemy's robot is defeated, time-limited concludes once a preset time duration runs out, and move-limited ends after a preset number of gem swaps. "Shot clock" mode initially gives participants three seconds to make each move, which decreases or increases by 0.1 seconds for each subsequent move based on the participant's respective success or failure in making a valid move. This is designed to induce stress as the participant is forced to make matches within progressively shorter response windows, eventually reaching equilibrium at approximately 50% accuracy with barely enough time to react. Failure to make a move on time results in the participant being unable to attack the enemy robot for one move, while the countdown to an enemy attack continues. Each hour of gameplay in a planned longitudinal study will be divided into three 15-minute rounds of normal, time-limited, and move-limited gameplay, each followed by a 5-minute session of "shot clock" gameplay.

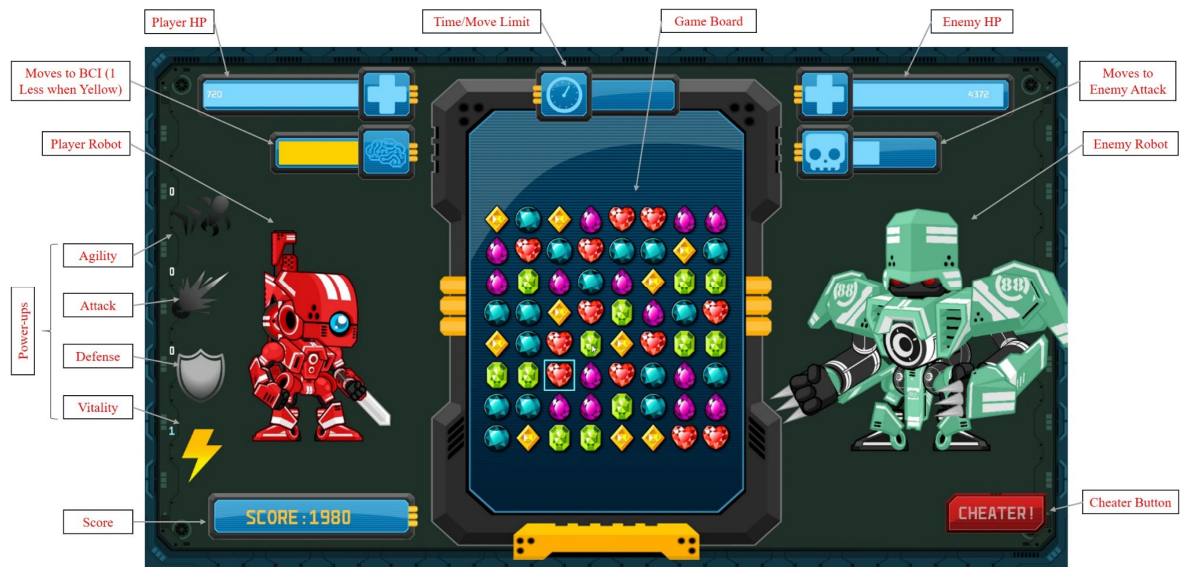

**Figure 1.** The tile-matching game interface consists of a central field of colored gems with the round timer above it, the participant's robot to the left, and the enemy robot to the right. Each robot has a health bar at the top of the screen, as well as bars indicating the number of gem swaps remaining before next BCI-relevant mini-game (left) and the next enemy robot attack (right). The left side of the screen also contains an inventory of earned Power-Ups (Agility, Attack, Defense, and Vitality), while the right side of the screen contains a "CHEATER" button for reporting gem-matching rule violations (see Error-Related Potentials, below).

The BCI mini-games incorporated here were designed to be minimally disruptive or even complementary to the flow of the game and sufficiently rewarding to encourage continued gameplay. The model chosen to incentivize engagement with the BCI tasks is similar to that found in common "free-to-play" (F2P) commercial games, which allow players to purchase power-ups to make gameplay easier. However, instead of participants being required to purchase upgrades with money, they play BCI-relevant mini-games and receive gameplay power-ups based upon their performance. To enable the regular collection of BCI data and to provide participants with regular power-ups, a randomized BCI mini-game is generally loaded after a preset number of participant-initiated gem matches, initially set to 5 moves. The power-ups which may be acquired through the BCI mini-games are: Attack (deal extra damage), Defense (reduces enemy attack damage), Agility (dodges one enemy attack), and Vitality (reduces the moves remaining until the next BCI-task/power-up opportunity). All power-ups are automatically used at the earliest appropriate time. Five BCI-relevant signals/tasks are the focus of the mini-games: rapid serial visual presentations (RSVP), steady state visually evoked potentials (SSVEP), motor imagery (MI)/motor execution (ME), error-related potentials (ErrP), and an n-back working memory task. An overview of each BCI paradigm's implementation in the game and analysis in the pilot follows.

## 2.1 Rapid Serial Visual Presentation (RSVP)

Rapid serial visual presentation (RSVP) tasks involve the participant searching for a target image in a series of more plentiful non-target images displayed at a relatively high frequency (5 Hz for this study). The RSVP task contains four levels of visual coherence, pseudo-randomly chosen at the start of each task to ensure an even distribution of coherence levels, ranging from clear presentation against a distinct background to images with heavily obfuscated borders between targets/non-targets and background. These were included in the game both to characterize individual behavior and EEG responses at each level and to create more variance and challenge to keep participants engaged with the task. At the beginning of each RSVP mini-game, the participant is assigned two target power-ups randomly from the possible four power-ups, shown on the left and right sides of the screen, and asked to squeeze a grasp switch on the respective side when a target image appears. The relatively rare targets (~13% frequency) are images of the power-ups and are embedded among non-target images of gems from the main game. A minimum target-to-target interval (TTI) of 0.8 seconds is enforced, but otherwise all images are randomly shuffled in the desired target/non-target ratio. Power-ups are rewarded based upon the percentage of targets correctly identified and number of false positives.

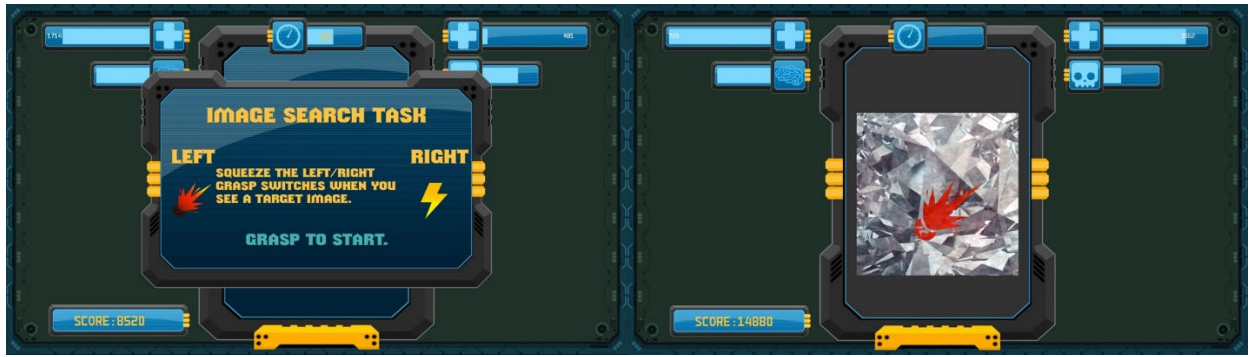

**Figure 2.** The interface of the RSVP task, with the instructions shown on the left and the task shown on the right. The participants watch a series of images flash by at 5Hz and when they see a target image, such as the "attack" (left) and "vitality" (right) power-ups shown in the task instructions in the figure, they squeeze a grasp switch on the respective side. In the example presented, the participant would squeeze the left grasp switch to respond to the "attack" power-up presented.

The primary signal associated with RSVP tasks is a P300 ERP, which is initiated when a subject observes an expected target image. This manifests as a positive power increase, most prominently localized in electrodes near the parietal lobe of the brain, approximately 300 ms after the stimulus. ERPLab was used to average across all trials of RSVP for all participants to visualize the P300. An additional check on the validity of the data was performed by averaging across trials of RSVP locked to non-targets, in which a clear SSVEP signal should appear at the presentation rate of the images (5 Hz).

## 2.2 Steady State Visually Evoked Potentials (SSVEP)

Steady state visually evoked potentials (SSVEP) are signals that arise when a subject views a flashing stimulus. They may be visualized as an increase in power of the EEG frequency band matching the frequency of the visual stimulus. Performing a power spectral density (PSD) analysis of EEG data from electrodes near the occipital lobe best captures these potentials. In the context of the game, this task is executed by placing a power-up in one of four boxes on the screen. The participant is asked to stare at the target box while it flashes at a set frequency, which may be 7, 9, 11, or 13 Hz. The participant is then told to click on the target as soon as the flashing ends, which is meant to incentivize the participant to focus on the target. The participant is rewarded with power-ups based upon their reaction time in clicking the target, with penalties if the participant anticipates and clicks too early.

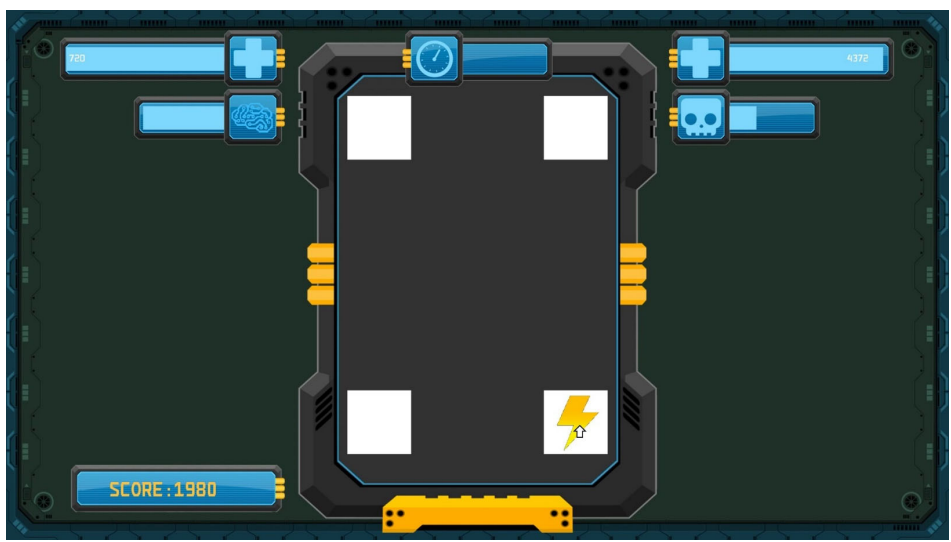

**Figure 3.** The interface for the SSVEP task is shown. The bottom-right hand corner contains a "vitality" power-up. The subject would stare at the target box while each box begins to flash at 7, 9, 11, or 13 Hz (pseudo-randomized). As soon as it stops, the participant is instructed to click on the target box. The white arrow is included to indicate the desired participant action.

## 2.3 Motor Imagery (MI)/Motor Execution (ME)

At the start of each MI/ME mini-game, a barrel appears in the center of the screen with an arrow underneath it that directs the participant to move the barrel to the left (to heal their robot) or to the right (to damage the enemy robot). The participant will then attempt to move it either left or right by imagining or executing movement with the respective hand. This task will alternate between motor execution (ME), which will require the participant to repeatedly squeeze a left or right grasp switch to move the barrel, and MI, in which the participant will imagine squeezing the grasp switch repeatedly and a classifier will determine from their EEG data which direction they are attempting to move the barrel. The arrow beneath the barrel will darken with successful execution of the task. After a 6-second period, the barrel will move in the desired direction if the participant has passed a pre-defined threshold of success, measured as either as a sufficient number of grasp switch squeezes per time or trials of successfully classified motor-imagery. The pilot study classifier has been trained from pooled EEG data collected in a similar task without feedback to discriminate between imagined left- and right-hand movements.

Classifier performance was used as the metric for analysis in the pilot data. This was done to ensure that the initial classifier had some ability to classify motor imagery signals in individuals not included in the training data. The signal generated by the individual will be analyzed in future papers, but as a non-reactive brain signal the profile of this signal in a single session is unlikely to reflect upon the design of the game and is thus beyond the scope of this paper.

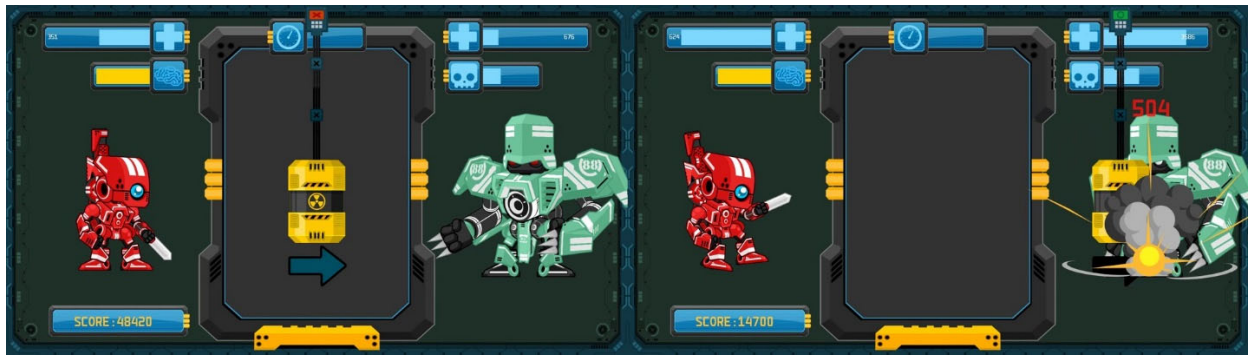

**Figure 4.** An example of a trial of the MI task requiring the participant to imagine repeatedly squeezing the right grasp switch. The left figure shows the active component of the task where classification of EEG is taking place, with the arrow below the barrel indicating both which grasp switch to imagine squeezing and the successful classification of the participant MI (darkening with increasing success). The right figure shows the successful outcome of the task (dealing damage to enemy), assuming that the participant produced a sufficiently distinct signal and the classifier was able to accurately label a sufficient number of windows as right-handed motor imagery.

## 2.4 Error-Related Potential (ErrP)

The error-related potential (ErrP) task is unique in that it occurs during standard gameplay. An ErrP is induced by making and recognizing a mistake or observing an outcome that is incongruent with one's action. During each successful gem match made by the player, there is a 15% chance that the game will substitute the participant's valid gem match with an invalid move, with a caveat that this cannot happen two moves in a row. If the participant notices that their valid move was substituted with an invalid move and not executed, they can report the "cheat" by clicking a button on the screen marked "CHEATER!" to execute their initial move and receive two power-ups. If the participant misses this erroneous swap, they may continue gameplay at no cost, though their initial gem match will not be completed. The ErrP was analyzed using a 1-second window after an erroneous gem swap occurred that was subsequently caught by the participant, as indicated by pressing the "Cheater" button.

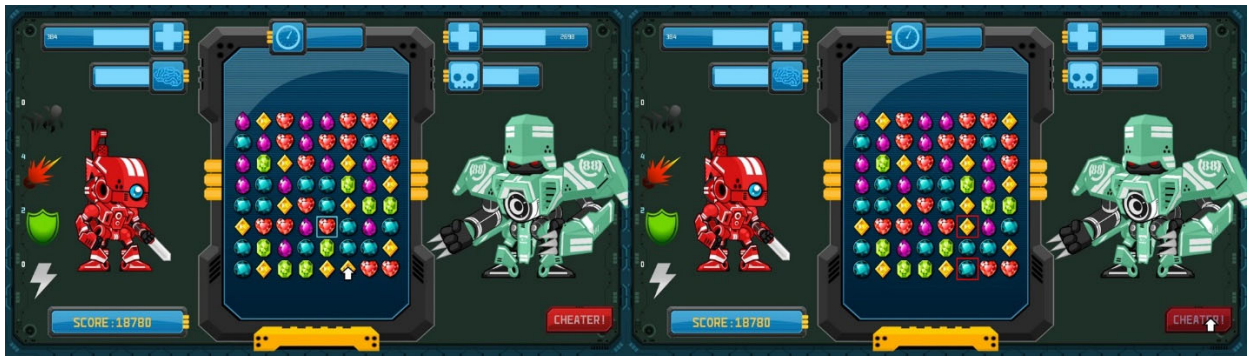

**Figure 5.** An example of the ErrP-inducing task appearing during gameplay. In the left figure, the player initiates a valid gem-match, which is then substituted for a random, invalid gem swap indicated by the red squares in the right figure. At this point, the participant would click the "CHEATER!" button to complete the initial match and earn power-ups. White arrows are added to each figure indicate the participant action at each step.

### 2.5 N-Back

The final mini-game included is an n-back working memory task, in which images of the power-ups are presented and the participant is asked to press the mouse button when an image matches an identical image presented a pre-set number of iterations (n) back. The images are presented for 0.8 seconds with an inter-stimulus interval (ISI) of 1.5 seconds. This task is executed with varying levels of n (n = 1, 2, 3, and 4) to induce varying levels of working memory load. Power-ups are rewarded based upon the percentage of targets identified and the number of false positives reported. Effective design of the n-back was analyzed by evaluating the performance profile across the four levels of difficulty and ensuring that the task evoked an expected P300 when the participant recognized a target stimulus.

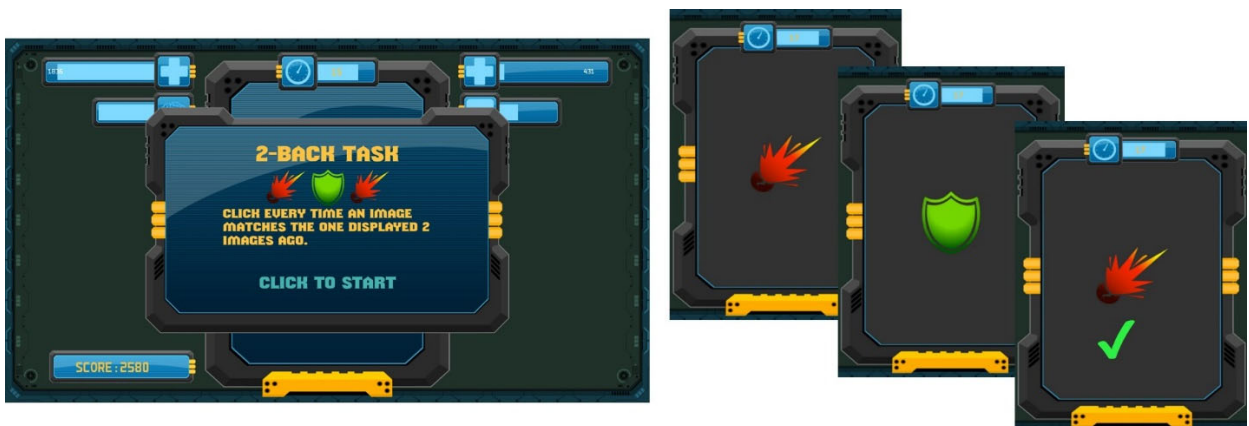

**Figure 6.** The instructions for this instance of the n-back working memory task inform the participant that it is a 2-back task. The sequence of images on the right hand indicate a correct reaction on a 2-back task, with the "attack" power-up with the green check mark matching another displayed two images previous. The green check mark is included to indicate when a participant should click and is not present in the actual game.

### 2.6 Pilot Study

A pilot study was conducted to determine the efficacy of the game as a research platform. Data were collected from 7 participants who each completed a single data collection session comprised of approximately 1 hour of BCI mini-game play (approximately 6 minutes of RSVP, 5 minutes of SSVEP, 12 minutes of n-back, and 5 minutes of ME/MI; no ErrP). The results of the pilot study provide validation that the BCI mini-games elicited neural signals indicative of those expected from more traditional BCI tasks, enabled the analysis of player performance to assess relative difficulty of the n-back task, and allowed a stress test of the game and data pipeline with real participants.

## 3. Results

All EEG data included in this analysis was bandpass-filtered between 1-40 Hz and subjected to independent component analysis (ICA), with the component most likely representing blinks removed from the dataset. This was the extent of generalized pre-processing and any further processing will be listed by signal/task. The standard error for each signal is included when relevant as a semi-transparent overlay upon the averaged EEG signal.

### 3.1 Rapid Serial Visual Presentation (RSVP)

Data from the RSVP task was analyzed in a 1-second window time-locked to the presentation of a non-target gem or a target power-up. All trials of correctly identified targets and non-target stimuli were averaged and baselined against the prior 200 ms to produce clear ERPs. Figure 7 shows the results of this analysis with each of the visual presentations marked. A P300 was observed in the target group, while a 5 Hz SSVEP signal with peaks approximately every 200 ms appears in both the target and non-target groups, though it is largely obscured in the target group by the P300. Subjects were asked to squeeze a grasp switch upon recognition of a target, but the average response time was ~500 ms, so it is unlikely that the apparent P300 is heavily influenced by EMG.

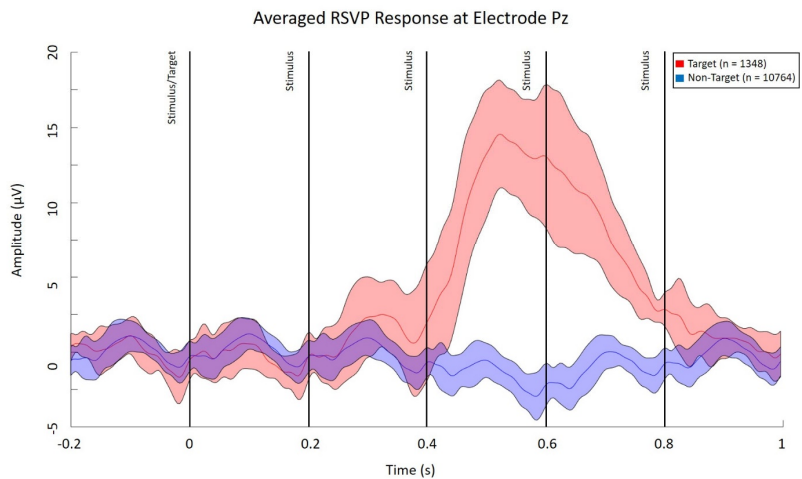

**Figure 7.** Averaged EEG response at electrode Pz, with standard error as a semi-transparent overlay, time-locked to the presentation of a target and non-target stimulus. The P300 is observed as a clear amplitude increase around 400 ms in the target group while the 5 Hz SSVEP response is observed in both groups as small increases in amplitude every ~200 ms, though it is largely obscured by the P300 response in the target group.

### 3.2 Steady State Visually Evoked Potentials (SSVEP)

All SSVEP data were separated into four groups representing each of the four target frequencies: 7, 9, 11, and 13 Hz. Each group had its EEG responses averaged and the PSD for each group was calculated using MATLAB. The PSD results for each subject were then averaged to create the following figure. All four targets had a peak at their expected frequency, however, there were also smaller peaks outside the expected ranges. This is hypothesized to be the result of the display of all four flashing boxes on the screen (potentially within the same visual field), participant behavior (drifting gaze or excessive blinking), and/or presentations being occasionally off-frequency due to imperfect synching with the monitor display rate (60 Hz).

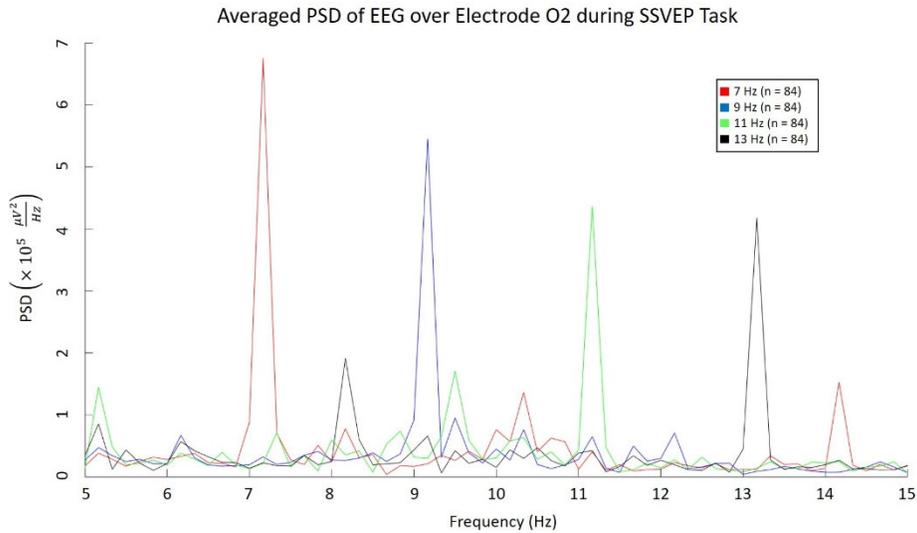

**Figure 8.** The averaged PSD of EEG from electrode O2 acquired from pilot participants during the SSVEP task. Significant power increases are observed at the proper frequencies for each frequency within the task: 7, 9, 11, and 13 Hz. Smaller peaks are visualized across the spectrum, potentially resulting from participant behavior, display of multiple flashing boxes on the same screen, and/or the display of occasional off-frequency stimuli. The peak at ~14 Hz averaged from the 7 Hz task is expected as a harmonic response.

### 3.3  Motor Imagery (MI)

The motor imagery task lacks a clear indicator of the performance of the game. While either classifier performance or representative topological EEG features could be analyzed, the classifier is expected to have low performance until sufficient data is collected to update the classifier using participant data, and without training it is unlikely that participants will generate consistent signals. As a general trend, performance on the left-hand task was substantially better than the right-hand performance, but this was observed to vary between subjects. This implies that the classifier is doing more than classifying randomly and that some subjects are able to consistently produce a recognizable MI signal without specific training. Increased performance is expected as the longitudinal study progresses, due to participant training and classifier updating.

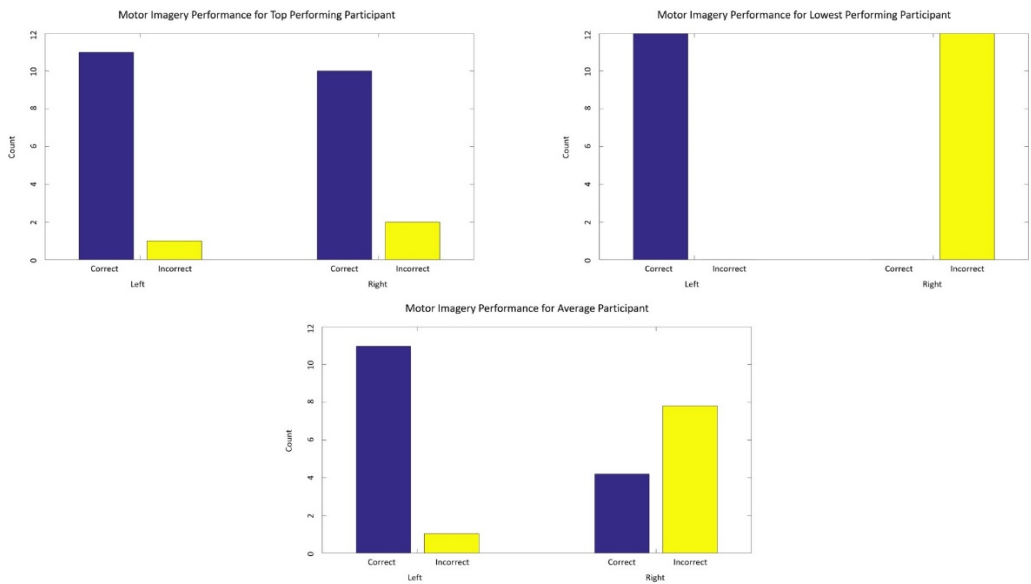

**Figure 9.** Performance the motor imagery task for the best (top left), worst (top right), and average participant (bottom). General results reveal a trend toward correct classification of left-hand motor imagery, though it's difficult to establish if this is the result of inconsistent signals from inexperienced participants, poor classification accuracy from the classifier, or more likely a blend of the two. The presence of high-performing individuals suggests that the classifier is not universally biased to predict a left-hand imagined movement.

### 3.4 Error-Related Potentials (ErrP)

The precise shape and timing of error-related potentials differ depending upon the event that triggers them, but a study with a similar subversion of user input to that in this game found that the ErrP consisted of a negative peak, likely feedback-related negativity (FRN), at 287 ms, followed by a positive potential at around 367 ms, and finally an N400 at around 486 ms (Spüler and Niethammer, 2015). The ErrP induced by the tile-matching game demonstrated an EEG waveform very similar to that described by Spüler & Niethammer, though the positive peak did not significantly differ from the non-target waveform.

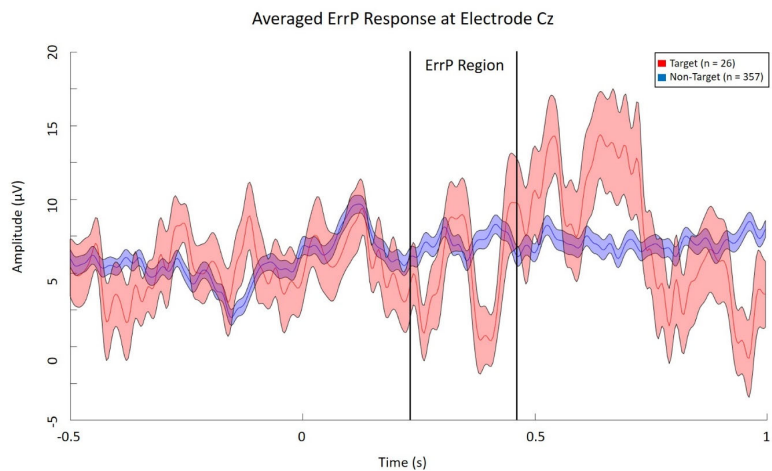

**Figure 10.** An averaged ErrP waveform acquired during technical testing of the game. The "Target" signal is time-locked to an erroneous gem swap, while the "Non-Target" is locked to a valid gem match. The "ErrP Region" shows an ERP with a waveform very similar to that expected from the literature. The two negative peaks of the response are significantly different from the non-target response, though the positive peak, while present, is not significantly defined from the non-target" response.

### 3.5 N-Back

Participants in the pilot study completed four trials of each level of difficulty with 5 targets per trial, for a total of 20 targets. One pilot subject did not complete the n-back task and was excluded from this analysis. The relatively sparse physiologic data from the pilot sessions limited analysis to ensuring that P300s were elicited when viewing targets and comparing performance between levels of difficulty. Direct analysis of workload will be conducted as part of the longitudinal study. A near-identical profile of the early P300 is observed in both groups, though they visually vary in the later portion of the ERP. This is likely due to the same visual stimuli being used for targets and non-targets in this task, varying only by context. The large standard error near the non-target P300 reflects high inter-individual variability in the difference between the target and non-target P300s, with some significantly distinct and others nearly identical.

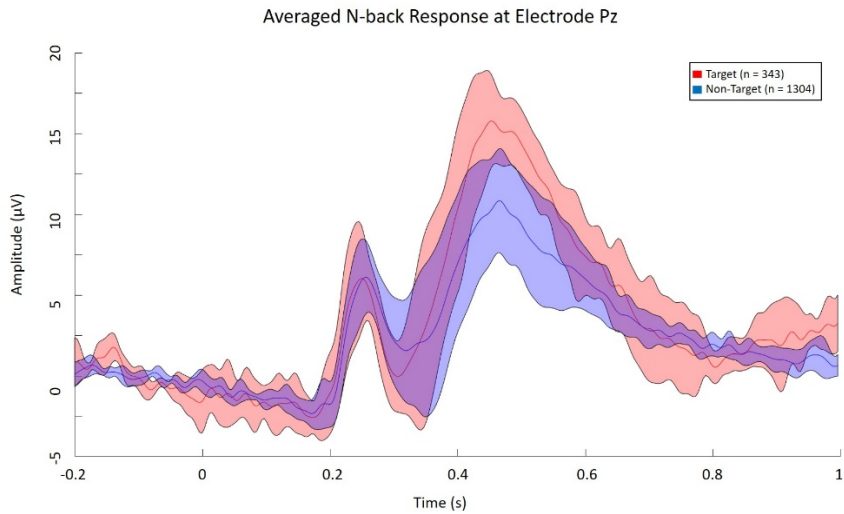

**Figure 11.** Averaged EEG over Pz time-locked to a recognized target or non-target. The responses show a very similar shape, with the target group reaching a higher amplitude than the non-target group. The near-identical early component is hypothesized to be due to the same visual stimuli being used throughout the task, while the late-stage difference reflects the recognition of the target/non-target status of the stimuli. While the averaged waveform is very similar, multiple participants had significantly distinct target and non-target late-stage P300 waveforms.

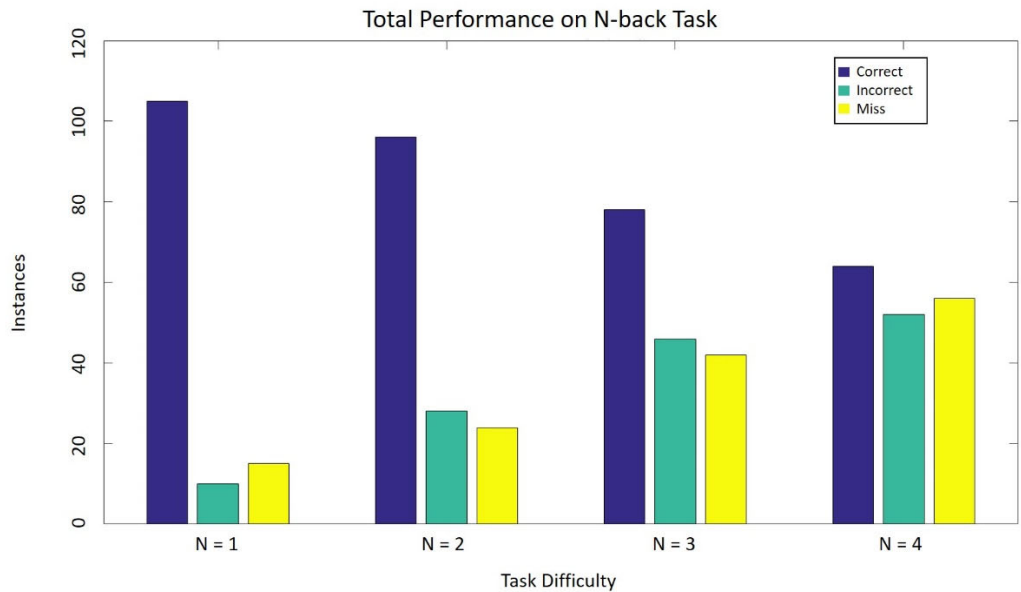

**Figure 12.** The performance profile for the various levels of difficulty of the n-back working memory task across all pilot subjects, demonstrating the expected decrease in performance with increasing difficulty. There are 20 targets at each difficulty level for a total of 120 possible targets. An incorrect response is when the participant identifies a non-target as a target due to error or guessing.

## 4. Discussion

The successful implementation of the "Game with a Purpose" model will enable large-scale, longitudinal data collection of behavioral, EEG, and physiological data from BCI-relevant tasks, which can be paired with evaluations of individual differences in user traits as well as assessment and/or experimental manipulation of various interdependent states (e.g., sleep pressure, stress, mood, time-on-task fatigue, cognitive load, etc.) and the interactions between them. Rich data of this nature may contribute directly to enabling creation of more robust BCI classifiers capable of improving the performance of both online, human-in-the-loop and *post hoc* BCI classification algorithms that function in real-world settings by leveraging knowledge about inter- and intra-individual changes in neural states. This robustness is a key challenge to fielding BCI technologies that are able to operate at high levels of performance outside of a laboratory setting, or even within the laboratory in contexts where high BCI literacy levels are essential. Longitudinal EEG and physiological data would also facilitate classifiers that are able to determine "brain states" directly, an important step toward extending group-model-based BCIs into the individualized, adaptive technologies that are so essential for future human-autonomy teaming capabilities (DeCostanza et al., 2018). For instance, EEG and/or physiological profiles associated with excessive workload could be leveraged to initiate interventions to reduce workload or to dynamically re-task other members of a team to assist with completion of an objective. For high-risk tasks involved in military operations or even everyday tasks such as driving, knowledge of these states in time to initiate proper interventions could increase task performance and potentially save lives.

As support for this concept, early testing of this game has proven it to be capable of eliciting the responses and behavior expected from the respective BCI-relevant tasks. Including these tasks as mini-games in the context of a larger game should provide participants with greater levels of reward, resulting in increased engagement that should facilitate longitudinal study of learning and natural variability in cognitive states over time (such as sleep pressure and stress). Additionally, inclusion of different game modes (e.g., time- or move-limited), variable session durations, and variable or even adaptive levels task difficulty (e.g., different levels of "n" in the n-back task, attenuated clock time in "shot clock" mode) allow for manipulation of states like workload and fatigue to further assess within-session EEG and physiological variability. This will enable quantification of natural and task-related variability in interdependent neural states, the interaction between them, and their impact on neural signals that cannot be captured using cross-sectional or short-duration research approaches, providing valuable insight into inter- and intra-individual differences in BCI literacy that have been regularly observed, but not yet explained. In today's age of crowdsourcing and the "Quantified Self," where physiological monitors and sleep trackers are built into watches, smartphone apps are capable of tracking subtle changes in day-to-day states ranging from mood to diet and everything in between, and even EEG systems that are becoming more affordable and realistic for personal, at-home use, games with a purpose of this nature have the potential to reach thousands or even eventually millions of users, making it more important than ever for novel research approaches of this nature to be developed now in order to improve and drive development of the technologies of the future.

5.  Acknowledgments